\documentstyle[preprint,prc,aps]{revtex}
\def\shiftleft#1{#1\llap{#1\hskip 0.04em}}
\def\shiftdown#1{#1\llap{\lower.04ex\hbox{#1}}}
\def\thick#1{\shiftdown{\shiftleft{#1}}}
\def\b#1{\thick{\hbox{$#1$}}}
\begin{document}
\title{The complete version of Moscow $NN$ potential}

\author{Vladimir I. Kukulin, V. N. Pomerantsev,}
\address{Institute of Nuclear Physics, Moscow State
University, 119899 Moscow, Russia}
\author{and Amand Faessler}
\address{Institute for Theoretical Physics,
University of T\"ubingen, Auf der Morgenstelle 14,\\
D-72076 T\"{u}bingen, Germany}

\maketitle

\begin{abstract}
A complete version of the Moscow $NN$ potential model is presented. The
excellent description for all essential partial waves has been found in the
energy range 0 -- 350 MeV. The one-channel version of the model includes the
orthogonality condition to most symmetric six-quark states in all lowest
partial waves and thus, from this point of view, the model generalizes the
well known Saito's orthogonality condition model (OCM) for the baryon-baryon
interaction case. The specific features of the presented model which
distinguish it from many conventional force models are discussed in
details. One of them is a specific tensor mixing between nodal and nodeless
wavefunctions which results in very reasonable values of the OPE cut-off
parameter $\Lambda =0.78$ GeV and the $\pi NN$-coupling constant
value $ f^2=0.075$ in nice agreement with modern trends. The model, in case
of its confirmation in precise few-nucleon calculations, can lead to
noticeable revisions for many nuclear properties given by  conventional
force models.
\end{abstract}

\section{Introduction}

In spite of the great progress attained recently in the construction of the
modern realistic $NN$-potentials of second generation, based on the concept
of improved one- and two-meson exchange~\cite{Nij93,Stoks93,Wir95,Lac81},
a large number of unsolved problems are still left
in the field. The majority of the problems here is related to description
of the short-range part of $NN$-interaction and to the quantitative
description of few-nucleon systems \cite{Wit91,Gron97,Autr98,Fons98,Sato98}.
In particular, one of the basic
problems is connected with the consistent incorporation of gluon-
and quark-exchange degrees of freedom and their "matching" with the
meson-exchange concept.

One of the most fundamental difficulties here is how to avoid
double counting of the same effects in the gluon- and
meson-exchange sectors of the unified interaction. Another basic problem
is the fact the true six-quark microscopic Hamiltonian is presently unknown.
While some
effective three-quark Hamiltonians which include chiral symmetry
breaking and confinement \cite{Gloz96,Isg78,Bir85}
have been developed to describe the  baryon spectra, there
is no guarantee that the same Hamiltonians can also be applied
for  six- (and multi) quark systems.

Recently \cite{Wit91,Gron97,Autr98,Fons98,Sato98} some fundamental problems
 have also been found in
the consistent description of few-nucleon and meson-few-nucleon systems.
On the one hand, the
numerous calculations made in recent years for quark-effects in various
few-nucleon observables~\cite{Bey83,Malt85} have shown, in general, quite
moderate contributions of such effects at low energies and momentum
transfers~\cite{Bey83,Malt85,Buch96,Buch89}.
On the other hand, however, quite remarkable disagreements between
the data and the most accurate three- and four-nucleon calculations
have been found \cite{Wit91,Gron97,Autr98,Fons98}. They probably can be
ascribed to an improper treatment of the quark degrees of freedom.
This follows from the fact that the above mentioned few-nucleon
calculations do include 3N-force and $\Delta $-isobar
effects together with the most realistic $NN$ interactions, i.e. they
include, to our current knowledge, all essential contributions.

Thus, developing a quantitative $NN$-interaction which
includes properly both quark- and meson exchange effects and is, on the other
hand, not so difficult to handle and complicated as those $NN$ models derived
from multiquark Hamiltonians, is still very topical. Besides, in view of our
insufficient knowledge of the accurate six-quark Hamiltonian, the theory must
be constructed in a manner to avoid features of QCD which presently can not
be handled with confidence, e.g. in particular the
$qq$ interaction, the form of confinement (e.g. linear or quadratic) etc.
At the same time, however, it is highly desirable to incorporate into the
interaction model some general reasoning about the preferred symmetry of
the six-quark system in various $NN$ channels, about the characteristic size
of six-quark states etc. In such a case the conclusions derived from the model
will result mainly from general symmetry requirements and general structure
of the model etc. rather than some particular choice of parameters for the
$qq$ interaction, or for the particular law of confinement etc.

We will demonstrate in the present work that only a few basic assumptions
are quite sufficient for the derivation of such a hybrid model. On this
basis quark effects in the few-nucleon physics can be described more
reliably. Our consideration is based on the assumption (which is common for
all hybrid models) that both nucleons merge somehow their quark contents at
short ranges into different six-quark states dependent on the partial wave,
the energy and the total spin. While at intermediate and large distances
where the nucleons do not overlap
noticeably with each other the interaction mechanism is governed by
one-meson exchange (which was just the original Yukawa idea about the origin
of strong interaction \cite{Yuk35,YukSak35}). This is a common basis of all
hybrid models and one model is distinguished from other one by the way of
matching of inner quark- and external meson-exchange channels. E.g. in
hybrid models due to Kisslinger~\cite{Kis82} and Simonov~\cite{Sim84}, the
matching of both channels is done at some arbitrarily chosen hypersphere
with radius $R$, although the matching conditions in both
models~\cite{Kis82,Sim84} are quite different.

In a sharp contrast to such hybrid models we prefer to do this matching
in the Hilbert space of the six-quark states with different symmetries, where
every such a state is constructed from single-particle harmonic
oscillator~(h.o) quark orbits \cite{Obu91,Oka83,Yam86}. Thus,
in accordance to this idea we subdivide a total
Hilbert space ${\cal H}$ of the six-quark states on two mutually orthogonal
subspaces ${\cal H}_P$ and ${\cal H}_Q$:
$${\cal H}={\cal H}_Q+{\cal H}_P.$$

In the first subspace ${\cal H}_Q$ we include six-quark states with highest
possible spatial symmetry, where one has a maximal overlap of the
single-quark orbitals. Whereas the six-quark states with a lower spatial
symmetry are placed in its orthogonal complementary subspace. The most
symmetrical states can be shown to have the structure
which is rather similar to compound states in a spherical or weakly deformed
bag. On the other hand the
states of lower symmetry include a few $p$-quark orbitals like $s^4p^2$,
$s^3p^3$ etc. and these mixed symmetry states, being projected out onto
$NN$-channel (of unexcited
nucleons), result into the nodal $NN$-radial wavefunctions \cite{Obu91}.
 Accordingly, their structure is analogous to clusterized peripheral states.

On the basis of above considerations and also of other arguments of
symmetry character we have suggested in the previous works
\cite{LIYAF86,Pom87,Kuk92,KuFaes98,Kuk98} a two-
component model for baryon-baryon interaction with two mutually orthogonal
channels. Then, by subsequent exclusion of six-quark compound states one
comes to an effective one-channel potential model of Moscow type, in which a
deep $NN$-potential well includes as its eigenstates the most symmetrical
six-quark states (to be more precise, their projections onto
the $NN$-channel).
As a result of combining two different components into one channel
for the effective interaction, the orthogonality condition between $NN$
scattering states and localized six-quark states in such a model is
satisfied automatically due to the hermiticity of the Hamiltonian. In a
consistent realization of such a program the wavefunctions of the
$NN$-relative motion
in the "external", i.e. clusterized channel are generally not
to be related to the six-quark wavefunctions in the "inner" channel. Moreover,
it is very likely they should be wavefunctions belonging to quite different
Hamiltonians. The underlying dynamics of the most symmetric six-quark
states must be very tightly interrelated to specific chromodynamic effects
such as quark- and gluon condensates, instantons, breaking chiral invariance
etc. whereas the external channel should be describable in terms of
meson-exchange.

Thus, these two different channels can hardly be described
consistently by a unified Hamiltonian (at least, on the up-to-date level of
our knowledge for low-energy QCD). The price to perform technically this
description is a different dynamics in the multiquark- and meson-exchange
channels\footnote
{It should be emphasized that in currently developed models of baryons in
which the $qq-$ interaction is described via one-meson
exchange~\cite{Gloz96,Obu91}
these one-meson degrees of freedom are nothing else but effective degrees of
freedom. Thus, these degrees of freedom  in multiquark system will be somehow
different from those in three-quark system.}. And hence it is highly
desirable to employ for these two components a two-channel model with
mutually orthogonal channels.

This reasoning justifies our model from the physical point of view.
Moreover, its real success in description of $NN$ partial amplitudes
demonstrated in the present paper allows to justify the model {\em a
posteriori}!

The structure of the work is as follows. In the Section II we present our
approach for the construction of the two-component hybrid model of
$NN$-interaction and
its interrelation to possible dibaryons. In the Section III we present
a realization of the generalized orthogonality-condition model (GOCM). In
Section IV the above GOCM is constructed and the structure of the
one-channel potential is discussed. Section V is devoted to
a quantitative description of $NN$-phase shifts in the energy range
0 -- 350 MeV. We give here also the effective-range parameters and a
detailed discussion of the structure of the deuteron. In Section VI
we discuss specific nonconventional interference and tensor mixing
between nodal and nodeless wavefunctions and the cut-off parameters for
the meson-exchange potentials at short range. And finally, the main results
of the work are summarized in the Section VII. In the Appendix we give the
formulas for our interaction model in the momentum representation.

\section{A hybrid model with orthogonal components}

The nucleon-nucleon interaction at large and intermediate distances is well
known and can be described by meson-exchange
potentials~\cite{Stoks93,Lac81}. The internal
nucleon degrees of freedom (quark and gluon ones, if we start from the quark
model for nucleon) do not show up in this approach. However, when the
nucleons come closer than $\sim $1 fm, a transition of the $NN$ system into
other channels arises, i.e. the internal nucleon
degrees of freedom begin to be of crucial importance. As a dynamic model,
e.g., a six-quark bag model can be used.

However we still have no full dynamic model describing all possible states
of the two-nucleon system. Therefore we divide the full Hilbert space
${\cal H}$ (including both nucleonic and non-nucleonic degrees of freedom)
into two orthogonal subspaces~\cite{LIYAF86,Pom87,Kuk92,KuFaes98,Kuk98}:
\begin{equation}
{\cal H}={\cal H}_{NN}\oplus {\cal H}_{6q}
\end{equation}
These subspaces must be orthogonal because the dynamics in them is
essentially different: one of them - ${\cal H}_{NN}$ - includes only nucleonic
degrees of freedom and meson - exchange dynamics, whereas the other,
named
${\cal H}_{6q}$, includes $6q$-model dynamics (or QCD-inspired dynamics).
Accordingly to (1) we introduce two mutually orthogonal projection operators
$P_{NN}$ and $P_{6q}$. It is important that the (unknown) full Hamiltonian of
the system does not commutate with  $P_{NN}$ and $P_{6q}$ and contains
transitions between the $NN$- and the $6q$-channels.

If we suppose the existence of a full Hamiltonian $H$ obeying the
6$q$ Schr\"{o}dinger equation:
$$H\psi =E\psi $$
one can easily obtain, following Feshbach \cite{Fes62}, the effective
Hamiltonian for $NN$-component:
\begin{equation}
\psi _{NN}\equiv P_{NN}\psi
\end{equation}

\begin{mathletters}
\begin{equation}
P_{NN}HP_{NN}\psi
_{NN}+P_{NN}HP_{6q}[P_{6q}(E-H)P_{6q}]^{-1}P_{6q}HP_{NN}\psi_{NN}=E\psi_{NN}
\end{equation}
\begin{equation}
P_{6q}\psi_{NN}=0
\end{equation}
\end{mathletters}

In accordance with these ideas the effective nucleon-nucleon Hamiltonian
$h_{NN}\equiv P_{NN}HP_{NN}$ includes only meson-exchange potentials:
\begin{equation}
h_{NN}=P_{NN}HP_{NN}=t+v^{ME}
\label{(4)}
\end{equation}
The second term in eq. (3a) is an effective potential which couples $NN$-
and $6q$-channels and will be designated further as $v_{NqN}$. With these
notations, the effective equation with orthogonality condition (3b) takes
the form:

\begin{mathletters}
\begin{equation}
(h_{NN}+v_{NqN})\psi_{NN}=E\psi_{NN}
\end{equation}
\begin{equation}
P_{6q}\psi_{NN}=0
\end{equation}
\end{mathletters}

As an effective wavefunction in $NN$-channel $\psi _{NN}$ one can naturally
use the resonating group ansatz (RGA):
$$\psi _{NN}={\cal A}(\varphi_{_N} \varphi_{_N} \tilde{\chi}_{_{NN}}),$$
in which $\varphi_N$ is the nucleon wavefunction and $\tilde{\chi}_{_{NN}}$ is the
wavefunction for the $NN$-relative motion obeying the orthogonality
constraint (3b).

The approach above formulated can be considered as a general formal
framework for the hybrid model of the $NN$ interaction. The parameters of
$v_{NqN}$ can be determined from the underlying six-quark
Hamiltonian~\cite{Obu91,Oka83}, if
one assumes that the effective $qq$ interaction is the same in $3q$- and
$6q$-systems, or by fitting the $NN$-scattering phase shifts. (A quite
similar procedure has been used, e.g., in the quark compound bag model
due to Simonov \cite{Sim84}.)

There are two essential differences in our approach from other hybrid
models. These are the orthogonality condition in eqs. (5) and
the structure of total Hamiltonian. It should be emphasized that equations
(3) and (5) are fundamentally different. Eqs. (3) is formally
derived from the full Schr\"{o}dinger equation by means of the identity
transformations. The orthogonality condition does not play a role
in eq. (3a) except at $E$=0. On the contrary, the eqs. (5) are  model
equations,
which don't involve the full Hamiltonian $H$. Therefore the presence of
the orthogonality condition (5b) is absolutely necessary.

\section{Two-component model in frame of constitueuent quark model
and Moscow potential}

To fill the general scheme (5) with a microscopic content it is necessary to
use some approximation for the full Hamiltonian $H$. This can be done in
the frame of the constituent quark model (see e.g. our previous paper
\cite{KuFaes98}). Symmetry considerations allowed to identify the
subspace ${\cal H}_{6q}$. It consists out of square integrable functions
$\psi_{6q}$ describing the lowest $6q$-bag states with maximal spatial
symmetry: $|s^6[6]>$ for $S$-waves and $|s^5p[51]>$ for $P$-waves.

This choice of ${\cal H}_{6q}$ can be justified by several
independent reasons \cite{Yam86,LIYAF86,Pom87,Kuk92,KuFaes98,Kuk98}.
E.g., recent chiral model calculations \cite{Stan97}
have shown that the structure of fully symmetric 6$q$ states $|s^6[6]>$, in
contrast to the mixed symmetry states as $|s^4p^2[42]>$, cannot be described
by the cluster RGM-ansatz and that they are quite similar to the shell-model
ground states
of magic nuclei. However, the most conclusive argument in favor of the
s
separation of 6$q$-states with high and low symmetry arises from our general
understanding of quantum chromodynamics, where the effective interactions
must essentially depend on number and type of quarks.
Moreover, if we assume,
that some effective bosonisation of initial QCD-degrees of freedom occurs in
the peripheral area of nucleon and thus this bosonic mode is an important
component of interquark interaction~\cite{Gloz96} one can conclude,
taking into account a highly nonlinear character of such bosonization, chiral
meson fields must play a crucial role in the dynamics of the six-quark configurations.
Such chiral fields should stabilize strongly these six-quark components of interaction.
Thus, by approximating
the $Q$-space Green function $[P_{6q}(E-H)P_{6q}]$ with one pole at
$E=E_{6q}$ one gets a separable form for the potential $v_{NqN}$
\begin{equation}
v_{NqN}=P_{NN}|H|\psi_{6q}>(E-E_{6q})^{-1}<\psi _{6q}|H|P_{NN}
\label{6}
\end{equation}
where $$E_{6q}=<\psi_{6q}|H|\psi_{6q}>$$

As a projection operator onto the $NN$-channel one can employ a respective
operator taken from the resonating group method (RGM):
\begin{equation}
P_{NN}={\cal A}|\psi_{N}\psi_{N}>{\cal N}^{-1}<\psi_{N}\psi_{N}|{\cal A}
\label{7}
\end{equation}
where $\psi_{N}$ is the three - quark function of the nucleon,
${\cal A}$ is the antisymmetrizer, and ${\cal N}$ is the overlap kernel:
\begin{equation}
{\cal N}= <\psi_{N}\psi_{N}|{\cal A}|\psi_{N}\psi_{N}>
\label{8}
\end{equation}
With this choice of the projection operator $P_{NN}$, the model equation (5)
becomes a two-body effective Schr\"{o}dinger equation for the orthogonalized
relative motion wave function $\tilde {\chi }(R)$

\begin{mathletters}
\begin{equation}
\left (T_R+V^{ME}+ 10\frac{|f><f|}{E-E_{6q}}\right )\tilde{\chi}=
E\tilde{\chi}
\end{equation}
\begin{equation}
<g|\tilde{\chi}>=0
\end{equation}
\end{mathletters}

in which
\begin{equation}
<{\bf R}|f> \equiv f({\bf R})=<\psi_{6q}|H|\psi_{N}\psi_{N}>
\label{10}
\end{equation}
\begin{equation}
<{\bf R}|g> \equiv g({\bf R})=<\psi_{6q}|\psi_{N}\psi_{N}>.
\label{11}
\end{equation}
In a good approximation one can take a delta-function for the
overlap kernel ${\cal N}({\bf R},{\bf R}')$  \cite{Obu91,Oka83}:
\begin{equation}
{\cal N}({\bf R},{\bf R}')\simeq \frac{1}{10}\delta ({\bf R}-{\bf R}').
\label{12}
\end{equation}

We emphasize once again that eq. (9a) with the orthogonality condition (9b)
is
not equivalent to the full six-quark Schr\"{o}dinger equation $H\psi=E\psi $.
Actually we suppose we know only individual parts of the full Hamiltonian:
\begin{enumerate}
\item[-] subHamiltonian $h_{NN}=t_{R}+v^{ME}$, acting in the subspace ${\cal
H}_{NN}$ and describing the meson-exchange interaction between unexcitable
nucleons, and
\item[-] other subHamiltonian $H_{6q}$, describing the lowest states in
$6q$-bag (in given case $H_{6q}=\sum E_{6q}|\psi_{6q}><\psi_{6q}|$).
\end{enumerate}
Thus the full six-quark Hamiltonian is needed only for determination of
coupling
between the subspaces in eq.(6). In this model the $6q$-bag functions
$\psi_{6q}$ are not eigenfunctions of the full Hamiltonian (otherwise
$[P_{6q},H]=0$ and $v_{NqN}\equiv 0$). Moreover, it is obvious that the sum
of the projectors $P_{NN}$ (7) and $P_{6q}=\sum |\psi_{6q}><\psi_{6q}|$ is not
unity in the full six-quark space ${\cal H}$. Therefore, eq. (9) cannot be
formally deduced from the full
Schr\"{o}dinger equation and the orthogonality condition (9b) proves to be
necessary.

The effective two-nucleon equation (9a) provides the basis for developing
the local and nonlocal parts of $NN$-interaction models of Moscow type. The
main point here is just the orthogonality condition (in $S$- and $P$-waves),
which results in appearance of nodes in $NN$-scattering wave functions, the
positions of the nodes being do not depend on energy (at least up to
laboratory energies $E_{NN}\sim$1 GeV). The term $v_{NqN}$ provides
an additional
attractive interaction at $E<E_{6q}$. It has been shown in previous papers
\cite{LIYAF86,Pom87,Kuk92,KuFaes98,Kuk98}, that the phase shifts and nodal
behavior of wave functions typical
for eq. (9) are well reproduced by a deep local attractive potential with an
extra bound state and the respective orthogonality condition constraint. So,
from this point of view, the $NN$-interaction model, known today as Moscow
potential, is the simplest local model which ensures the orthogonality
between the scattering wave functions and the most symmetric $6q$ states
$|s^6[6]>$ projected onto  the $NN$-channel. However the situation for
$P$-waves turns out to be already different. Attempts to achieve a
satisfactory description of the phase shifts by using a local attractive
potential failed for these partial waves \cite{Pom87}. Therefore, one needs
to use the general orthogonality condition model (GOCM) presented here.

\section{Structure of the potential}
\label{Struc}

Here we give the full version of the $NN$ {\em potential} model with
the additional orthogonality condition in $S$- and $P$-waves. The potential
is {\em an effective one-component approximation} to the two-component model,
described in the previous section. Actually we have replaced the nonlocal
term $V_{NqN}$ (attractive at low energies) in eq.~(9) by an additional local
attractive well.

The total interaction is however highly nonlocal due to the presence of the
$S$- and $P$-wave projection operators which are employed in order to take
into account the orthogonality condition (9b). As a result we
do not require locality, this means we have a weaker interrelation
between  the orthogonality condition and the form of  the attractive well.
This
decoupling of the attractive potential from the orthogonality condition
improves essentially the approach. In particular, the quality of the
fits for $P$-waves gets more accurate than in the old-fashioned Moscow
model with eigenprojection\cite {LIYAF86,Pom87,Kuk92}. Besides the
matrix eigenstate projector in coupled $^3S_1-{ }^3D_1$ channels, as was
demonstrated in our previous paper~\cite{KuFaes98}, can be replaced quite
accurately by a scalar one-channel projector.

In order to use a potential with the orthogonality conditions in few-body
calculations, one has to add the projection operator with a very large
positive coupling constant to the local part of the potential, in all $S$-
and $P$ partial waves~\cite{KuFaes98}.

For the sake of uniformity and convenience we include similar separable
terms, but with finite coupling constants, also in some other partial waves
($D$ and $F$). These terms replace the standard spin-orbital part of
the interaction (for even-parity waves) and reduce partially a strong
attraction due to the central part of  the local potential. In fact,
these separable terms imitate a short-range repulsion generated by
$\omega$-meson
exchange\footnote{It should be emphasized here that the $\omega $-exchange
terms in traditional meson-exchange models
are highly nonlocal due to form factors and energy- and momentum
dependence.}. We include also the tensor interaction which couples
partial waves with angular momenta $l$ and $l\pm 2$. It can be quite
accurately described by a truncated OPE-potential in all partial
waves with the channel coupling being determined by a truncation parameter.

In the present version of Moscow potential we have replaced the Gaussian
form of the central potential which have been used in all previous versions
of the model\cite{LIYAF86,Pom87,Kuk92,Kuk85} by an exponential one. We have
found the exponential form gives
a more satisfactory description of  the phase shifts, in particularly for
 the $^3S_1 - { }^3D_1$-channel (see also the refs.~\cite{Dub97}).

Thus, the model potential consists out of three parts:
\begin{equation}
v_{NN}=v^{loc}_M + v^{OPE} + v^{\rm sep}
\label{13}
\end{equation}
 where the local exponent well $v^{loc}_M$ depends on the channel spin and
 parity:
\begin{equation}
v^{loc}_M(r)=V_0 \exp (-\beta r)+({\bf s}{\bf l})V^{ls}_0\exp(-\beta_1r).
\label{14}
\end{equation}
In the state-dependent  separable part
\begin{equation}
v^{\rm sep} = \lambda|\varphi >< \varphi|
\label{15}
\end{equation}
a Gaussian form factor $<r|\varphi>=\varphi (r)$ is used:

\begin{equation}
\varphi(r)=N r^{l+1} \exp \left ( -{1 \over 2} \left ( {r \over r_0} \right
)^2 \right )\label{16}
\end{equation}
with normalization condition $\int\varphi^2 dr=1$. The integer $l$ labels
the partial waves.

For  the one-pion-exchange part of the potential  the standard
form with a {\em dipole} form factor is chosen:
\begin{equation}
v^{OPE}({\bf k}) = \frac{f^2_{\pi}}{m} \frac{1}{{\bf k}^2+m^2} \left (\frac
{\Lambda^2-m^2}{\Lambda^2+{\bf k}^2}\right )^2 ({\b\sigma}_1{\bf
k})({\b\sigma}_2{\bf k}) \frac{({\b\tau}_1 {\b\tau}_2)}{3}
\label{17}
\end{equation}

With such a form factor choice the OPE tensor potential vanishes at
the origin as
it should. In the coordinate representation the OPE-potential has the form:

\begin{equation}
v^{OPE}(r) = \frac{({\b\tau}_1 {\b\tau}_2)}{3}\frac{f^2_{\pi}}{4\pi} m \left
(f_C(r) ({\b\sigma}_1{\b\sigma}_2)+ f_T (r) \hat S_{12} \right )
\label{18}
\end{equation}
where the tensor operator
\begin{equation}
\hat S_{12} = \frac{({\b\sigma}_1{\bf r})({\b\sigma}_2{\bf r})}{r^2}-
\frac{({\b\sigma}_1{\b\sigma}_2)}{3};
\label{19}
\end{equation}
and
\begin{equation}
f_C(r) = (\exp(-x)-\exp(-\alpha x))/x- (\alpha^2-1)\alpha /2
\exp(-\alpha x);
\label{20}
\end{equation}

$$
f_T(r)= \exp(-x)/x(1+3/x+3/x^2) - \alpha^3 \exp(-\alpha x)/(\alpha x)
(1 + 3/(\alpha x) + 3/(\alpha x)^2)
$$
\begin{equation}
 -(\alpha^2-1)\alpha /2\exp(-\alpha x)
(1+1/(\alpha x));
\label{21}
\end{equation}
\begin{equation}
x=mr;\qquad \alpha=\Lambda/m.
\label{22}
\end{equation}

We use here  the averaged pion mass $m=(m_{\pi_0}+2m_{\pi_{\pm}})/3$  and the
averaged value of pion-nucleon coupling constant $f^2_{\pi}/(4\pi ) = 0.075$
as we don't wish to deal with the difference between
$np$ and $pp$ isovector phase shifts in the present work.

Thus only three free parameters $V_0$, $\beta$, and $\alpha$ are left for
the local part of interaction for each combination of spin and parity in
addition to two parameters $r_0$ and $\lambda$ of the separable term in each
channel. It should be noted that only some of the values $r_0$ and $\lambda$
are independent free parameters (for $D$- and $F$-waves). Values of
$\lambda$ for $S$- and $P$-waves must go to infinity (in real
calculations the value of $\lambda \sim 10^5 -10^6$ MeV is quite enough).
Values for $r_0$ for these channels are related to the local attractive well
(for the local potential, the requirement of the best approximation for
eigen bound state by Gaussian (16) defines $r_0$ uniquely).
Thus, we have
totally 32 parameters of the potential (and the value of $\pi NN$  coupling
constant) giving a very good description of all $N-N$ partial waves (except
of some high $l$ channels) in the wide energy range $0-400$ MeV. The number
of parameters almost coincides with that for most recent version of the
Nijmegen $N-N$ potential~\cite{Stoks93}.
The parameters
for the present version of our $NN$ potential are given in Tables~I-II.

\section{Description of phase shifts and deuteron structure}

The potential parameters as given in Tables I-II were determined by fitting
the Nijmegen phase shifts (PWA93) \cite{Nij93}. In Figs.~1-3 the
recent SAID phase shifts (SP97) \cite{Arn92} are also
presented for comparison. As can be seen from the Figures, some
discrepancy between the results of both phase shift analyses (PSA) exist,
especially for some partial phase shifts. With applications to
few-nucleon problems in mind we tried to
reproduce with maximal accuracy the $^1S_0$ and $^3S_1- { }^3D_1$ phase shifts
and the values of the scattering length and  the effective range.

\subsection{Singlet partial wave channels}

The description of singlet $n-p$ phase shifts for both even- and odd
parities is illustrated on Fig.1. It is evident from the Figure that
the quality of fit to the data of recent phase shift analysis is quite
good, especially for the Nijmegen PSA-results. E.g. the fits in $^1S_0$
and $^1P_1$ channels are almost perfect.  The quality of fits can be
estimated quantitatively from the Table IV for these channels. The average
deviations for all singlet channels are only $0.1-0.2 \%$ excepting the
$^1G_4$-channel where the discrepancy with PSA-data is  largest and around
$1\%$.

Also there is a problem in precise description of the singlet effective
range $r_0$ (see the Table III).  We used as "experimental" the value $r_0$
presented in compilation of Dumbrajs et all of 1983 (see footnote to the
Table III). However,  in view of the very good agreement of our phase shifts
with the Nijmegen PSA for $^1S_0$-channel one could conclude that the
disagreement for $r_0$ should be really much reduced.

\subsection{The even-parity waves}

The $S$-wave potential turns out, as is in the previous versions of Moscow
potential \cite{KuFaes98}, to be strongly attractive. The Gaussian (15)
with the
range parameter $r_0$, included in the orthogonality condition (9b), is
close to  the eigenfunction of  the ground "forbidden" state in the potential.
In other words, we obtain for $S$-waves actually almost a local potential.
However, for the $^3S_1-{ }^3D_1$ channel (and also for all triplet coupled
channels) we use, strictly speaking, non-eigenstate
one-channel projector, as in \cite{KuFaes98}, in order to avoid a more
complicated two-channel eigenprojector.

We do not introduce here a spin-orbital potential for even partial waves
in an explicit form because it cannot be determined by PSA-data for
$^3S_1-{ }^3D_1$ channel, and the role of spin-orbital potential for
higher even-parity partial waves is played by the term $v^{\rm sep}$.

It should be kept in mind here that the complete two-channel version of our
model includes in the proper $NN$-channel one-meson exchange interaction
terms
(in a subspace orthogonal to symmetric six-quark compound states). Thus, in
the two-channel model, the spin-orbit terms should be described by a
conventional meson-exchange model. However in the effective {\em
one-channel} model presented here the separable state-dependent spin-orbit
interaction in even-parity channels is inavoidable to compensate partially
the strong attractive potential in  the $S$-wave.

The effective range parameters for singlet and triplet $S$-wave
channels are given in Table~III. Among all the calculated phase shifts the
maximal disagreement with PWA93 (though not large) is observed for the
tensor mixing parameter $\varepsilon _1$.

\subsection{The triplet odd-parity waves}

In the accordance to the microscopic quark picture, the orthogonality to
the bag-like functions $|s^5p [51]>$  must be
included for all $P$-waves. However, if we shall look at the behaviour of
"experimental" $P$-wave phase shifts at energies up to 500 MeV we shall
not find any
repulsion in  the $^3P_2$ channel, because the corresponding phase shifts are
purely positive until the energies $\sim $1 GeV. There is no repulsive
core for this channel also in  the majority of  the conventional
realistic $NN$ potentials.
But if we look to the phase shifts at high energies (see Fig.~4) one
can observe a repulsion appearing in all three triplet $P$-waves,
while $^3P_2$-phase shifts becoming negative only at energies higher 1~GeV.
From the point of view of the constraints imposed by the orthogonality
condition,
this means the function to which the $^3P_2$ scattering function is
orthogonal is much more narrow than that for other $P$-wave channels, $^3P_0$
and $^3P_1$.

It is interesting that fitting  the  $^3P_2$-wave at energies up to 350 MeV
enables us already to determine the range parameter $r_0$ of the projector
(see Table II). The inclusion of the projector improves appreciably
the description of  the phase shifts up to 350 MeV. An attempt to reproduce
the $^3P_2$-phase shift
using a purely attractive potential with "extra" bound state results in a
very deep ($\sim $15 GeV) potential and an unsatisfactory quality of the
description. Besides, such a deep potential is not suitable for description
of  the $^3P_0$ and
$^3P_1$ phase shifts. That is why we have waived in the present
version from the concept of a local Moscow model for $P$-waves.

So, for odd partial waves we have a rather small attractive well
($\sim $220~MeV)
and orthogonality to the non-eigen bound states for local part of potential.
It might mean the size of six-quark bag in $P$-waves should be smaller than
in $^3S_1$ and $^1S_0$ waves.
One notices here that the range parameters of the projectors for  the
$^3P_0$ and $^3P_1$-channels ($r_0\approx 0.32$ fm) are almost
coinciding with each other. As is seen from Fig.~4, attraction for some
odd higher partial waves with $L=J$ ($^3F_3$ and $^3H_5$) is noticeably
deficient in  the given version of the model.

Unlike the even partial waves, we used a more conventional local spin-orbit
potential for odd partial waves (see eq.(~14)) because the usage of the
separable spin-orbital form is not convenient to describe the splitting
of $P$-phase shifts.

It would be rather instructive to estimate the averaged relative difference
of phase shifts predicted by the Moscow model and the recent
phase shift analysis [1] using the criterion of {\em relative} difference
or the respective absolute difference measured in radians:

\begin{equation}
\varepsilon _{rel}=\frac{1}{N}
\sum_{i=1}^N \left| \frac{\delta^{pot}_{JSl,i} -
\delta^{PSA}_{JSl,i}}{\delta^{PSA}_{JSl,i}} \right| ^2
\label{23a}
\end{equation}

\begin{equation}
\chi ^2_{JSl}=\frac{1}{N}\sum ^N_{i=1} \left|
\delta ^{pot}_{JSl,i}-\delta ^{PSA}_{JSl,i} \right| ^2,
\label{23b}
\end{equation}
where $\delta ^{pot}_{JSl,i}$ and $\delta ^{PSA}_{JSl,i}$ are partial phase shifts
in the channels $JSl$ at the energy $E_i$ for the Moscow model and Nijmegen
phase shift analysis
respectively. The Table~IV presents the values of $\varepsilon _{rel}$ and
 $\chi ^2_{JSl}$ for all
considered $JSl$-channels. It is evident from the Table the average
deviation of phase shifts predicted by the Moscow model and recent PSA
is very small and around $0.2-0.4\%$. It means that the description of
$NN$-observables with the presented force model should be very good.

\subsection{Deuteron structure}
The accurate description of  the deuteron structure offers an additional
strong test for
any nuclear force model. Many deuteron properties,
even in the static limit, depend sensitively on the behaviour of the
$NN$ force at intermediate and short ranges~\cite{Buch96}, especially on the
$D$-wave contribution. For example, with the first version of the present
force model~\cite{Kuk85}, we found an impressive agreement
with experimental data for all crucial $D$-wave deuteron observables like
$Q_d$, $A_S$, $A_D/A_S$. But this early model included a node not only in
 the $S$-wave but also in the  $D$-wave. This extra node in $D$-wave was
a consequence of a very short-range truncation of  the OPE tensor
force~\cite{Kuk85} which contradicts somehow  the microscopic picture of
the underlying interactions (e.g. according to the wide-spread
opinion~\cite{Eric} the OPE tensor force cannot penetrate deeply inside the
two-nucleon overlap region).

Hence, in subsequent versions of the model~\cite{Kuk92,KuFaes98,Kuk98}, a
more soft cut-off factor has been employed which resulted in the disappearance
of the $D$-wave node. As an immediate consequence of the softer truncation in
 the OPE tensor force the $D$-wave deuteron observables have become close
to the values predicted by conventional force models, i.e. the values of
$\eta$ and $Q_d$ are a little bit underestimated (see Table~V).
Nevertheless the node in  the $S$-wave and  the strong attractive $S$-wave
potential,
tightly related to this, results in a very specific interference between
$S$- and $D$-wave components and a specific character of tensor mixing (see
Section~VI).

The values of deuteron observables for three versions of our force model are
presented in Table~V while the pattern of the deuteron wave functions is
displayed on Fig.~5. One can see on the Figure the short-range maximum
in the $D$-wave almost
disappears for the current version of the force model while this maximum
in the  $S$-wave gets rather
reduced. It is interesting to note the $D$-wave amplitude in  the
current version
of the model (solid line) is a little bit lower than in the previous versions
(dashed and dot-dashed lines) due to a smaller value of the derivative of
the $D$-wave component near the $S$-wave node ($\sim 0.53$ fm). While
the asymptotic behaviour of  the $S$-wave looks almost perfect (see values
of $A_S$ in Table~V).

Thus we can conclude from the deuteron results presented in this Section
that the short-range part of the tensor force needs to be a bit improved.
Careful inspection of the Table~V shows
unambiguously  the general good agreement for the deuteron parameters found
with the sharply different force models such as Nijmegen and Moscow
potentials. The values for the deuteron observables
are a result of some general properties (like OPE tail) and of the
$NN$ phase-shifts used for fitting only and essentially do not depend on the
details of the force at short ranges\footnote{Certainly this
conclusion may be invalid for non-static, e.g. energy-dependent or
multicomponent force models.}.

\section{ Specific interference between tensor and central forces and
the $\pi$NN coupling constant in the Moscow force model}

We included this specialized Section to the present work in order to
emphasize a specific character of interference between tensor and central
forces in the Moscow force model. This interference will be shown below to be
very advantageous in some aspects as compared to the traditional force models.
The main difference between our and traditional models  as concerned to
wave function form is the
nodal character of  the $S$-wave deuteron and scattering wavefunctions and
the practically nodeless character of the $D$-wave functions.\footnote{The
very
small inner maximum in the $D$-state wavefunction in the present version
(see the solid lines on the Figs.~6a-6b) can be ignored in
any calculation if we do not consider very high momentum transfer.} We will
show here that the specific tensor mixing between the $S$-wave state with a node and
the almost nodeless $D$-wave state results in a remarkably different
$\varepsilon_1$-behavior. We compare the $D$-wave observables with the
results of traditional models.

First of all we emphasize here that the best fit for $NN$ phase shifts is
attained in our case with a very reasonable value for  the OPE cut-off parameter
$\Lambda_{dip}=0.78$ GeV (we used here the dipole form factor), 
see eq.(17). This soft cut-off parameter is in
nice agreement with both experimental results
and with all theoretical
estimations made in $\pi$-N dynamics\cite{Arn92,Eric,Yudin,Mac87,Louc94}.
It should be contrasted
with a statement formulated in~\cite[p.232]{Mac89} for traditional
OBE-force model: "...a value of 1.3 GeV is {\em a lower limit} for
$\Lambda_{\pi}$". The conventional OBEP model with
$\Lambda=0.78$ GeV gives the extremely low values for $Q_d=0.238$ fm$^2$,
the ratio
$D/S=0.0233$ and $P_D=2.4\%$ \cite{Mac89} which should be compared to the
respective values for our force model (see Table~V).

      In despite of the "soft" value of $\Lambda$, the
$D$-wave deuteron properties in our model (see Table~V in Sect.~V) are in a
rather good agreement with the experimental data, being remarkably better
than the respective predictions of the traditional force models with the
same $\Lambda$ value. We note, in passing, that the harder truncation with
$\Lambda \simeq 1.3 \div 1.7$ GeV is usually taken in the
traditional force model just in order to fit reasonably the deuteron
properties and the tensor mixing parameter (see below).

        The second important point in the story is related to the
mixing parameter $\varepsilon_1$. In fact, in order to reach a reasonable
agreement with the recent phase shift analysis data for the
$\varepsilon_1$-mixing parameter \cite{Nij93,Arn92} the $\Lambda$-value must
be taken also around $1.5 \div 1.7$ GeV\cite{Mac89} (see Fig.~6) while the
same agreement with the experimental $\varepsilon_1$ is reached in our model
using a much more soft $\Lambda = 0.78$ GeV. This sharp difference from the
traditional force models can be ascribed to a different character of
mixing between $S$- and $D$-waves in our model.

Some additional confirmation comes from the value of $\pi NN$-coupling
constant
obtained in our model.  We choose here the Nijmegen force model as a good
representative of the traditional $NN$ potentials (see, the Table~V).
Two above models include practically the same values for the $\pi NN$
(charged) coupling constants ($f^2_{\pi NN}=0.075$ in our case\footnote{The
value corresponds to the charged coupling constant because we considered
first of all the $pn$ scattering phase shifts.} and $f^2_{\pi^{\pm}
NN}=0.0748$ for Nijmegen potential). The latter fact is very important
because the $D$-wave characteristics are directly related to the $\pi$NN
coupling constant. In this respect our model appears to corroborate the
smaller value of $g^2_{\pi NN} \simeq 13.60 $ advocated by the Nijmegen
group~\cite{Nij93,Stoks93}.

The two nice features of our model discussed above, i.e. the soft cut-off
parameter $\Lambda$ and low value of $\pi NN$ coupling constant, are in
agreement with modern trends and lend strong support to our model.

\section{Conclusion}
\label{conclusion}

The force model presented in this paper  differs in a few important
aspects from traditional $NN$ interaction models currently in use.
First of all the Moscow two-component model includes two mutually orthogonal
quark- and meson-exchange channels. This channel orthogonality leads to many
differences from the traditional force models. In particular it requires
a node in low partial waves with the node position almost independent
on the relative energy in a wide energy range ($\le 1$ GeV). The nodal behaviour of
wave functions is also preserved for the one-channel model presented here.
The node in  the $NN$ wave functions results in an enhancement of high
momentum components and a strong increase of
the average kinetic energy in  the deuteron and in  all few-nucleon
systems.  This increase of  the inner kinetic energy leads to significant
enhancement of higher angular momentum components in nuclei
and nuclear matter and also for many particular nuclear processes
\cite{KuFaes98,Kuk98} like $\pi$-meson absorption and scattering in  the
$\Delta$-resonance region etc.
This strong enhancement of high-momentum components in $N-N$ system as compared to any
traditional $N-N$ force model may be seen e.g. in hard bremsstrahlung process
$pp \to pp\gamma$~\cite{Neudd} at $E_p=300$ MeV and higher at small forward and backward
angles $\theta _{\gamma}$ of $\gamma$-emission. To make the comparison with 
traditional
repulsive core models most unambiguously the authors~\cite{Neudd} did their
bremsstrahlung calculations with both the Moscow model (in its previous 
version~\cite{Kuk92}) and its exact phase-shift equivalent supersymmetrical 
partner. Thus, such a comparison removes any questions on the possible 
on-shell origin of disagreements observed.

Redistribution of higher partial waves along Jacoby coordinates leads e.g.
to a noticeable enhancement of  the $P$-wave attraction for
$N+d$ and $N+2\alpha$ systems\cite{KuFaes98,Vor95}. The long-standing
puzzle of the analyzing power $A_y$ in low energy $N+d$ scattering is
explained by insufficient
attraction in just the $N-d$ relative motion $P$-wave~\cite{Wit91,Gron97}.
The apparent discrepancies for $n+{}^3$H elastic and $n+{}^3$He $\to d+d$
rearrangement low-energy scattering observed recently\cite{Fons98} appear
to have to be explained also by insufficient attraction in the
$n+{}^3$H(${}^3$He) $P$-wave\cite{Autr98,Fons98}.
Such enhancement of higher  partial wave contributions to  near-threshold-
and low-energy processes in few-nucleon and few-cluster physics when
replacing the deep Moscow-type potential (including extra bound states)
with its SUSY-partner potential - which is exact phase-shift equivalent -
is a sequence of some very general algebraic properties  of kinetic
energy operator in different coordinate systems and is disconnected at all to
any small variations in the on-shell properties of various $N-N$ potential
models of current use.

The second crucial point in the development of Moscow $NN$ force model is
the important role of the  six-quark components with maximal possible symmetry.
We showed recently that the coupling of the  meson-exchange
NN channel to the six-quark component can be strong enough to represent
adequately
the intermediate-range $NN$ attraction. In turn, this fact leads to quite
remarkable contributions of such six-quark configurations in nuclear bound
and low-excited states. If so, it may require some strong revision for many
nuclear properties as given by traditional force models (e.g. the
meson-exchange current contributions). Thus the strongest test for the new
model may offer few-nucleon calculations for  the analyzing power $A_y$ in  the
$n+d$ and $p+d$ low-energy scattering, for the analyzing power  $A_y$ in
$p+d$ radiative capture
reaction and for  the $p+d$ intermediate energy elastic scattering cross
sections (so called the Sagara puzzle\cite{Gron97,Autr98}).
Hence the careful
comparison of the predictions for few-nucleon systems using the Moscow
force and
more traditional $NN$ interactions may be extremely interesting.

\centerline{\large \bf Acknowledgments.}

We are thankful to many our colleagues for fruitful discussions
on the topics of the present study, especially to Profs. Steven Moszkowski 
and ~V.~G.~Neudatchin,
Dr.~A.~Buchmann and Dr.~I.~T.~Obukhovsky. We are also grateful to
Dr.~S.~Dubovichenko for careful checking our deuteron calculations.
The Russian authors thank the
Russian Foundation for Fundamental Research (grant No.97-02-17265) and the
Deutsche Forschungsgemeinschaft (grant No. Fa-67/20-1) for partial financial
support.

\newpage

\appendix
\section*{Moscow potential in momentum space}
The $K$-matrix defined as
\begin{equation}
2i Mq\hat{K}=\frac{1+\hat{S}}{1-\hat{S}}
\end{equation}
( $M$ is the reduced mass while $q$ is a linear momentum)
obeys the partial-wave Lippmann-Schwinger equation:
\begin{equation}
\hat K(q',q)=\hat V(q',q) + \frac{2}{\pi}\,{\rm P}\!\int k^2 dk
\frac{\hat V(q',k)\, \hat K(k,q)}{E-k^2/2M}
\end{equation}
where P  means the principal value integral.
The elements of matrix $\hat V$ in eq.(2) are equal to the partial-wave
momentum-space potential in $lSJ$ basis (up to factor $1/4\pi$):
\begin{equation}
V_{l'l}(q',q)=\frac{1}{4\pi}<(l'S)J|V({\bf q}-{\bf q'})|(lS)J>,
\end{equation}
where $V({\bf q})$ is related to $V({\bf r})$ by a standard Fourier
transformation:
\begin{equation}
V({\bf q})= \int e^{-i({\bf qr})} V({\bf r}) d {\bf r}
\end{equation}
        Here we give explicit formulas for all terms of the present version
of Moscow potential $V_{l'l}(q',q)$ in momentum space (in MeV$^{-2}$).

\subsection{Local part of Moscow potential $V_{ll'}^{loc}$}
\begin{equation}
V_{ll'}^{loc} = \delta_{ll'}\left\{\frac{V_0\tilde{\beta}}{2(qq')^2}
F_l(\frac{q^2+q'^2+\tilde{\beta}^2}{2qq'})
+[J(J+1)-l(l+1)-3/4]\frac{V_0^{ls}\tilde{\beta}_1}{2(qq')^2}
F_l(\frac{q^2+q'^2+\tilde{\beta}_1^2}{2qq'})\right\}
\end{equation}
where the parameters $\tilde{\beta}$ and $\tilde{\beta}_1$ are given in MeV:
$$
\tilde{\beta} = \beta \cdot \hbar c,  \qquad
\tilde{\beta}_1 = \beta_1 \cdot \hbar c  \qquad .
$$
$F_l$ is being the derivative of the second kind Legendre function:
\begin{equation}
F_l(x)=-\frac{d}{dx}Q_l(x);\qquad Q_l(x)=\frac{1}{2}\int_{-1}^1
\frac{dz\,P_l(z)}{x-z}
\end{equation}

\subsection{Separable terms of the potential}
In momentum space the separable terms with Gaussian form factors (15,16) have
the same form as in the coordinate space:
\begin{equation}
V_{ll'}^{\rm sep}(q,q')=\delta_{ll'}\,\lambda \,\frac{\pi}{2}\,
\varphi_l(q)\varphi_l(q'),
\end{equation}
where
\begin{equation}
\varphi_l(q)=
\left(\frac{2^{l+2}}{(2l+1)!!\sqrt{\pi}}\tilde{r}_0^{2l+3}\right)^{1/2}
q^l\exp (-\frac{q^2\tilde{r}_0^2}{2}).
\end{equation}
Here the normalization condition $\int \varphi_l^2(q)q^2dq =1 $ is assumed
and the factor $\pi /2$ is related
to the integration  measure used in eq.(A2), and $\tilde{r}_0$ is given
in  MeV$^{-1}$:
$$ \tilde{r}_0=r_0/(\hbar c).$$

\subsection{ The OPE potential with dipole truncation}
For the sake of reader's convenience we give also the known formulas
for OPE matrix elements.
\begin{itemize}
\item[a)] The central part of OPE potential:
\begin{equation}
({V_c^{OPE}})_{ll'}(q,q')=\delta_{ll'}\frac{(\bbox{\tau}_1\bbox{\tau}_2)}{3}
(\bbox{\sigma}_1\bbox{\sigma}_2)\frac{f^2_{\pi}}{4\pi}\frac{1}{2qq'}
\left\{Q_l(x)-Q_l(y)-\frac{\Lambda^2}{m_{\pi}^2}(y-x)F_l(y)\right\}.
\end{equation}
Here and below
\begin{equation}
 x=\frac{q^2+q'^2+m_{\pi}^2}{2qq'}, \qquad
 y=\frac{q^2+q'^2+\Lambda^2}{2qq'}.
\end{equation}

\item[b)] The tensor part of OPE potential for triplet uncoupled
channels with $l=J$:
\begin{equation}
({V_{\rm ten}^{OPE}})_{JJ}(q,q')=
\frac{(\bbox{\tau}_1\bbox{\tau}_2)}{3}\frac{f^2_{\pi}}{4\pi}\frac{1}{m^2_{\pi}}
\left\{\frac{q^2+q'^2}{qq'}G_J-\frac{2J+3}{2J+1}G_{J-1}-
\frac{2J-1}{2J+1}G_{J+1}\right\}
\end{equation}
where the function $G_l$ is introduced as follows:
\begin{equation}
G_l(q,q') = Q_l(x)-Q_l(y)-(y-x)F_l(y)
\end{equation}
and $x$ and $y$ are defined by eq.(A10)
\item[c)] The tensor part of OPE potential for coupled channels with
$l=J\pm 1$:
\begin{equation}
({V_{\rm ten}^{OPE}})_{J-1,J-1}(q,q')=
\frac{(\bbox{\tau}_1\bbox{\tau}_2)}{3}
\frac{f^2_{\pi}}{4\pi}\frac{1}{m^2_{\pi}}
\frac{J-1}{2J+1} \left\{\frac{q^2+q'^2}{qq'}G_{J-1}-\frac{2J+1}{2J-1}G_{J-2}-
\frac{2J-3}{2J-1}G_{J}\right\}
\end{equation}
\begin{equation}
({V_{\rm ten}^{OPE}})_{J+1,J+1}(q,q')=
\frac{(\bbox{\tau}_1\bbox{\tau}_2)}{3}
\frac{f^2_{\pi}}{4\pi}\frac{1}{m^2_{\pi}}
\frac{J+2}{2J+1} \left\{\frac{q^2+q'^2}{qq'}G_{J+1}-\frac{2J+5}{2J+3}G_{J}-
\frac{2J+1}{2J+3}G_{J+2}\right\}
\end{equation}
\begin{equation}
({V_{\rm ten}^{OPE}})_{J-1,J+1}(q,q')=
\frac{(\bbox{\tau}_1\bbox{\tau}_2)}{3}
\frac{f^2_{\pi}}{4\pi}\frac{3}{m^2_{\pi}}
\frac{\sqrt{J(J+1)}}{2J+1} \left\{2G_{J}-\frac{q'}{q}G_{J-1}-
\frac{q}{q'}G_{J+1}\right\}
\end{equation}
\begin{equation}
({V_{\rm ten}^{OPE}})_{J+1,J-1}(q,q')=({V_{\rm ten}^{OPE}})_{J-1,J+1}(q',q)
\end{equation}
\end{itemize}

\newpage

\begin{table}
\caption{Parameters of local part of the potential}
\medskip
\begin{tabular}{ccccc}
spin    & singlet & singlet & triplet & triplet  \\
parity  & even   & odd     & even    & odd \\
\hline
$\alpha$&6.08671 & 6.08671 & 6.08671 & 4.3160 \\
$V_0$   &-4346.19&-1767.26 &-4567.12 & -223.63 \\
$\beta$ &3.49366 & 2.84152 &3.81272 & 2.4959  \\
$V_0^{ls}$   &&&& -591.1 \\
$\beta_1$    &&&& 3.4688 \\
\end{tabular}
\end{table}

\begin{table}
\caption{Parameters of projectors and separable parts of the potential}
\medskip
\begin{tabular}{ccc}
State  & $\lambda$, MeV & $r_0$, fm \\ \hline
$^1S_0$&  $\infty$ & 0.3943 \\
$^1P_1$&  $\infty$ & 0.5550 \\
$^1D_2$&  107.2    & 0.4527 \\
$^1F_3$&  182.6    & 0.5191 \\ \hline
$^3S_1$&  $\infty$ & 0.3737 \\
$^3D_2$&  161.2    & 0.4695 \\
$^3D_3$&  588.2    & 0.3572 \\
$^3G_4$&   2.74    & 0.8077 \\
$^3P_0$&  $\infty$ & 0.3209 \\
$^3P_1$&  $\infty$ & 0.3226 \\
$^3P_2$&  $\infty$ & 0.1632 \\
$^3F_4$&   5.447   & 0.6221 \\
\end{tabular}
\end{table}

\begin{table}
\caption{Effective-range parameters for the potential variant given
in Tables 1 - 2}
\medskip
\begin{tabular}{ccccc}
    & \multicolumn{2}{c}{$a$, fm} & \multicolumn{2}{c}{$r_0$, fm} \\
    & theory & experiment & theory & experiment \\ \hline
triplet $^3S_1$& 5.422 & 5.419(7)\tablenotemark[1] & 1.754 & 1.754(8)\tablenotemark[1]  \\
singlet $^1S_0$& 23.74 & -23.748(10)\tablenotemark[2] & 2.66 & 
2.75(5)\tablenotemark[2]  \\
\end{tabular}

\tablenotetext[1] {S. Klarsfeld, J. Martorell, and D.W.I. Sprung, J.Phys.
{\bf G}: Nucl.Phys. {\bf 10}, 165 (1984) }

\tablenotetext[2] {O. Dumbrajs et al, Nucl.Phys. {\bf B216}, 277 (1983)}

\end{table}

\begin{table}[h]
\caption{Accuracy of fitting of phase shifts}
\medskip
\begin{tabular}{lcccccccccccc}
channel            &   $^1S_0$ &   $^1P_1$ &   $^1D_2$ &   $^1F_3$ &   $^1G_4$ &   $^1H_5$ &   $^3S_1$ &   $^3D_1$ \\
$\varepsilon_{rel}$\tablenotemark[1]
                   &0.007504 &0.000527 &0.002113 &0.000197 &0.012030 &0.001802 &0.004524 &0.000867 \\
$\chi^2$ per point\tablenotemark[2]
                   &0.005282 &0.001731 &0.000627 &0.000008 &0.000200 &0.000055 &0.005595 &0.003922 \\ \hline
channel            & $\varepsilon_1$&   $^3D_2$ &   $^3D_3$ &   $^3G_3$ &$\varepsilon_3$ &   $^3G_4$ &   $^3P_0$ &   $^3P_1$ \\
$\varepsilon_{rel}$
                   &0.006816 &0.000022 &0.034310 &0.027256 &0.023088 &0.001451 &0.000184 &0.000455 \\
$\chi^2$ per point
                   &0.000843 &0.000007 &0.000856 &0.004545 &0.013887 &0.000029 &0.000186 &0.003596 \\ \hline
channel            &   $^3P_2$ &   $^3F_2$ &$\varepsilon_2$ &   $^3F_3$ &   $^3F_4$ &   $^3H_4$ &$\varepsilon_4$ &   $^3H_5$ \\
$\varepsilon_{rel}$  &0.003874 &0.010679 &0.007830 &0.017762 &0.007590 &0.021424 &0.010617 &0.026970  \\
$\chi^2$ per point &0.018245 &0.000123 &0.000693 &0.001413 &0.000279 &0.000004 &0.000042 &0.000155  \\
\end{tabular}

\tablenotetext[1] {$\varepsilon _{rel}=\frac {1}{N}
\sum_{k=1}^N (\frac{\delta^{pot}_{JSl,k} -
\delta^{PSA}_{JSl,k}}{\delta^{PSA}_{JSl,k}})^{^2}$}

\tablenotetext[2] {$\frac{1}{N}\sum_{k=1}^N (\delta^{pot}_{JSl,k} -
\delta^{PSA}_{JSl,k})^{^2}$ (in radians)  }

\end{table}

\begin{table}[h]
\caption{Deuteron parameters for conventional and Moscow $NN$ potentials}
\medskip
\begin{tabular}{lcccccccc}
model    & $E_d$ (MeV)& $P_D$ (\%)& $r_m$ (fm) & $Q_d$ (fm$^2$) &
$\mu_d\,(\mu_N)$ & $A_S$ (fm$^{-1/2}$)& $D/S$
&$D_{\rm loop}$\tablenotemark[1]\\ \hline
RSC\tablenotemark[2]
       & 2.22461  & 6.47 & 1.957 & 0.2796 & 0.8429 & 0.8773 & 0.0262 & \\
Nijm 93      & 2.224575  & 5.754 & 1.966 & 0.2706 & 0.8429 & 0.8844 &
0.02524 & \\
Moscow 86\tablenotemark[2]
 &2.22444  & 6.57 & 1.966 & 0.2862 & 0.8422 & 0.8838 & 0.0268 &0.53\\
Moscow 98\tablenotemark[2]\tablenotemark[4]
 &2.22440  & 5.75 & 1.954 & 0.2708 & 0.8470 & 0.8746 & 0.0259 &0.30\\
present\tablenotemark[3]
 &2.22456  & 5.65 & 1.967 & 0.2731 & 0.8476 & 0.8845 & 0.0255 &0.08\\
experiment&2.224575(9) &      & 1.9660(68)& 0.2859(3) & 0.857406(1) &
0.8846(16) & 0.0256(1)\tablenotemark[5] & \\
\end{tabular}

\tablenotetext[1]  {$D_{\rm loop}$ is the relative amplitude of the $D$-wave
maxima, i.e the absolute value of the ratio of the first and second maximum
of the deuteron $D$-component.}
\tablenotetext[2]  {The value $\hbar^2/2m=41.47$ MeV$\cdot$fm$^2$ has been
used ($m=938.978$ MeV).}
\tablenotetext[3]  {The value $\hbar^2/2m=41.47107$ MeV$\cdot$fm$^2$ is used
($m=938.918$ MeV).}
\tablenotetext[4]  {Unfortunately, in our previous work~\cite{KuFaes98}
only rounded values for potential parameters are given in the Table~III.
The deuteron parameters cited in~\cite{KuFaes98} (for
variant B) do not correspond to the rounded potential parameters cited in
Table~III of ref.~\cite{KuFaes98}.
We thank Dr. S.B.~Dubovichenko who has attracted our
attention to this disagreement and give here the exact values for variant (B)
of Ref. [27]:
$V_O=-1329.18$ MeV, $\eta =2.2959$ fm$^{-2}$, $\alpha =1.8835$ fm$^{-1}$.}
\tablenotetext[5]  {The present value is taken from \cite{Rod90}}
\end{table}

\newpage

\centerline{\large \bf Figure captions}

\bigskip

{\bf Fig.1.} The comparison of spin-singlet phase shifts for the present
version of Moscow $NN$ potential with the data of the recent
energy-dependent phase-shift analyses: PWA93 \cite{Nij93} (circles) and
SAID97  \cite{Arn92} (triangles).

\bigskip

{\bf Fig.2.} The spin-triplet even-parity phase shifts for the present version of Moscow $NN$
potential. The data of the energy-dependent phase-shift analyses are:
PWA93 \cite{Nij93} (circles) and SAID97  \cite{Arn92} (triangles).

\bigskip

{\bf Fig.3.} The spin-triplet odd-parity phase shifts
for the present version of Moscow $NN$ potential. The data of the
energy-dependent phase-shift analyses are: PWA93 \cite{Nij93} (circles) and
SAID97  \cite{Arn92} (triangles).

\bigskip

{\bf Fig.3.} ({\em Continued.})

\bigskip

{\bf Fig.4.} The spin-triplet $P$-wave phase shifts in a wider energy region:
the data of energy-dependent phase-shift analysis SAID97 \cite{Arn92}
(solid lines) and predictions for present version of Moscow $NN$ potential
(dashed lines).

\bigskip

{\bf Fig.5a.} The deuteron S-wave and D-wave functions for
present (solid lines) and previous versions (\cite{Pom87} -- dashed lines
and \cite[variant B]{KuFaes98} -- dot-dashed lines)
of the $NN$ Moscow-type potential. The deuteron wave functions calculated
with the RSC potential (dotted lines) are shown for comparison.

{\bf Fig.5b} Short-distance zoom of Fig.~5a (see caption to the
Fig.~5a).

\bigskip

{\bf Fig.6.} The energy dependence of the mixing parameter $\varepsilon_1$
for different values of cut-off parameter $\Lambda$ corresponding to
conventional (dashed lines) and present (solid lines) force models.

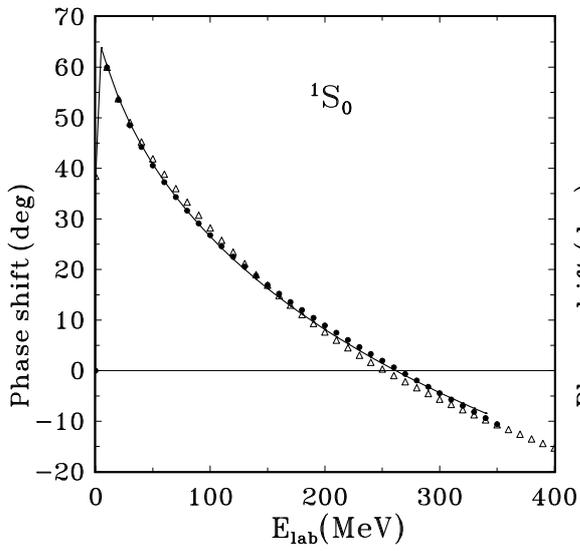
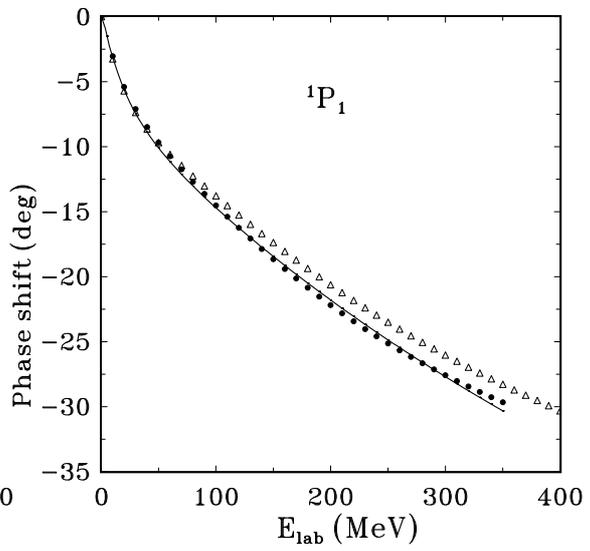
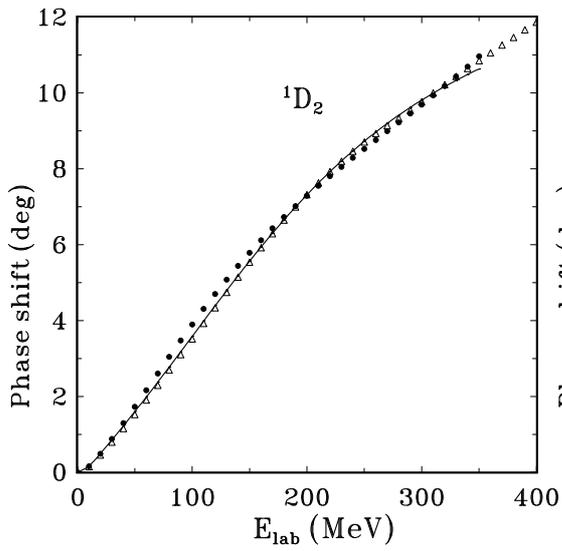
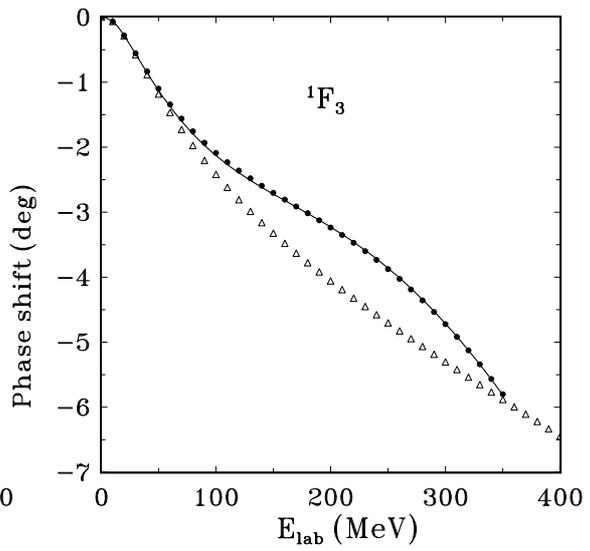
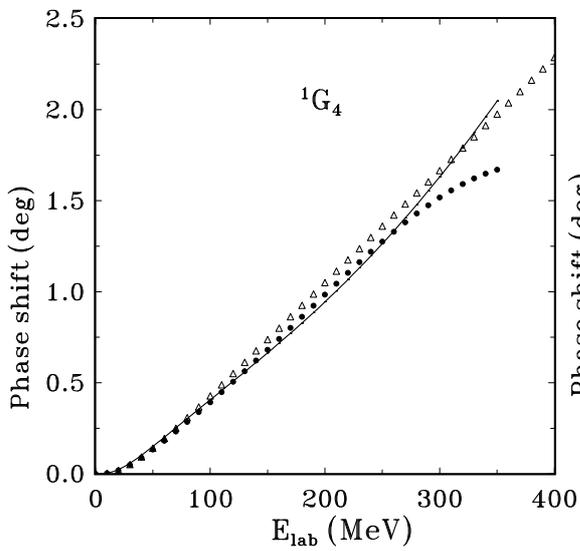
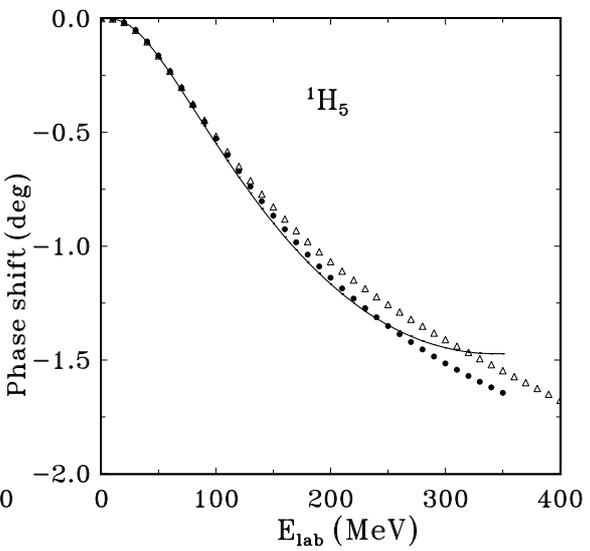
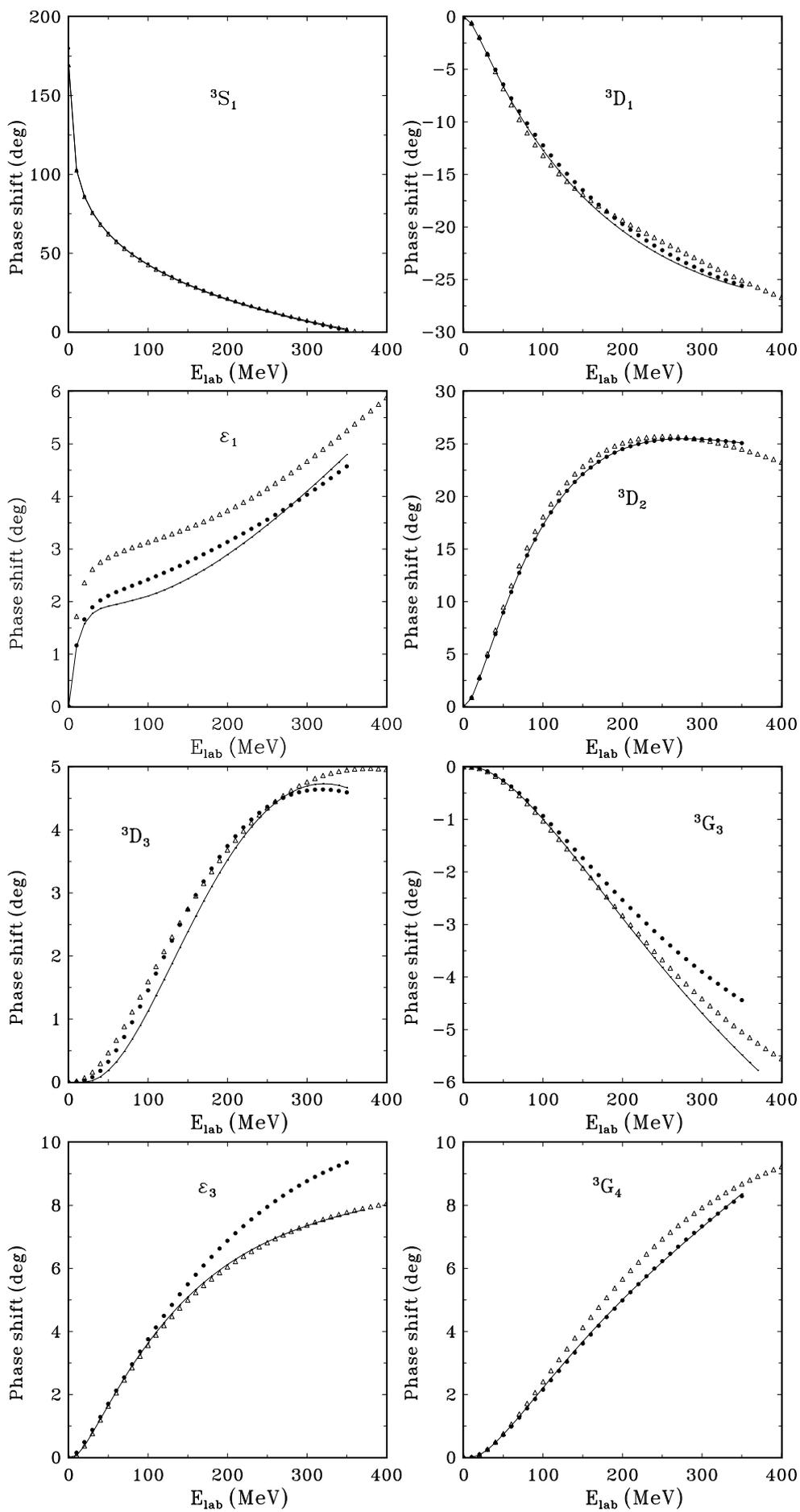
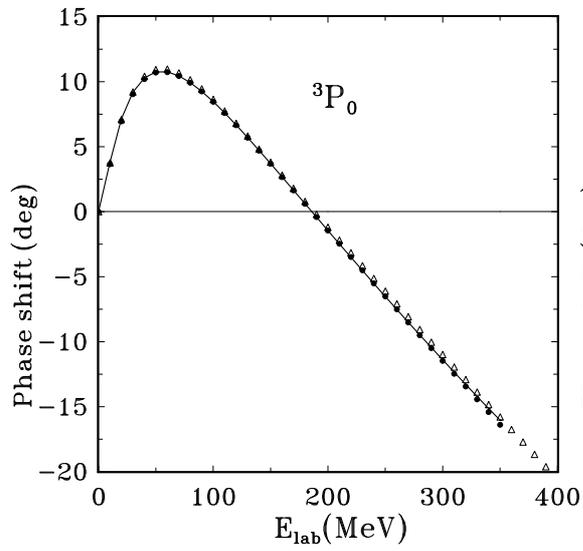
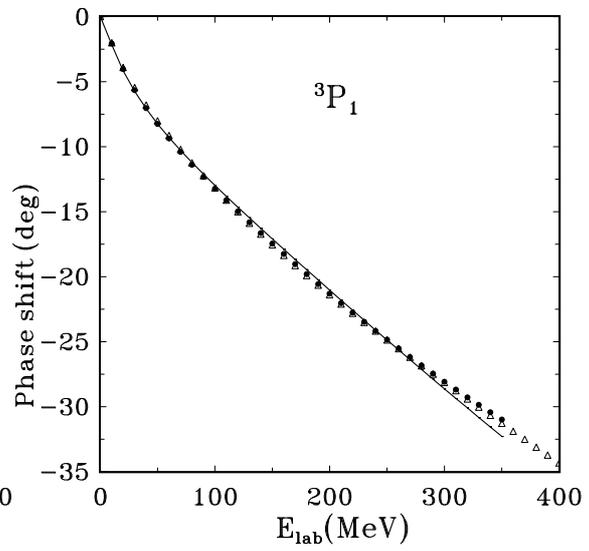
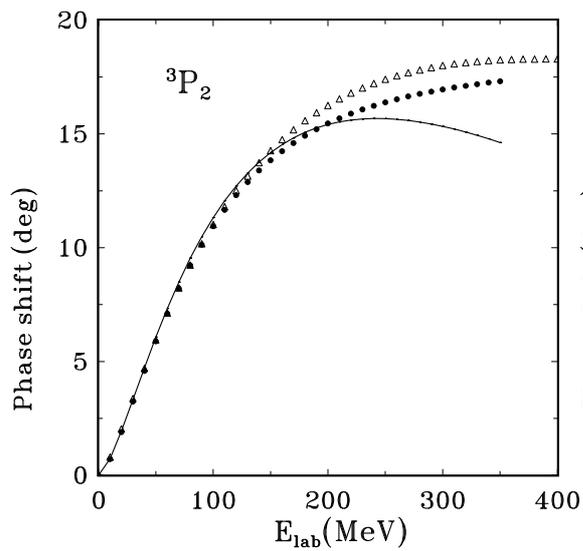
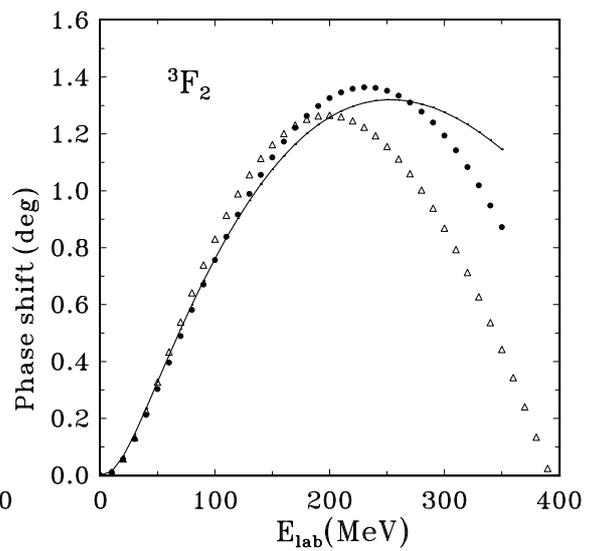
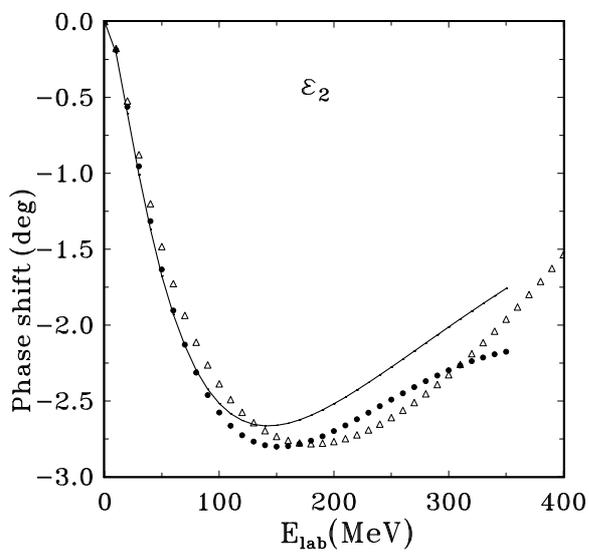
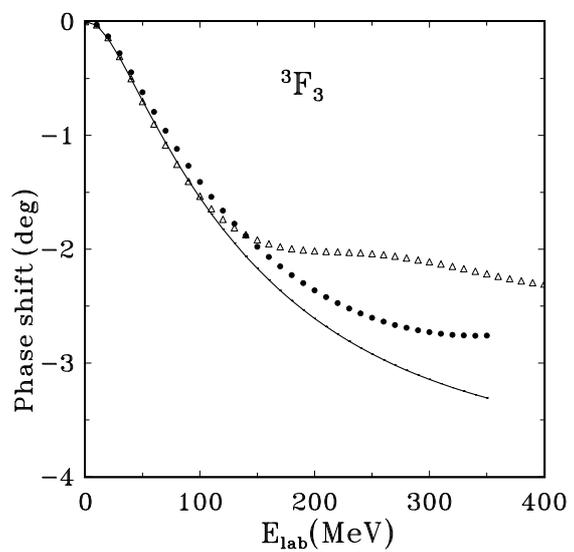
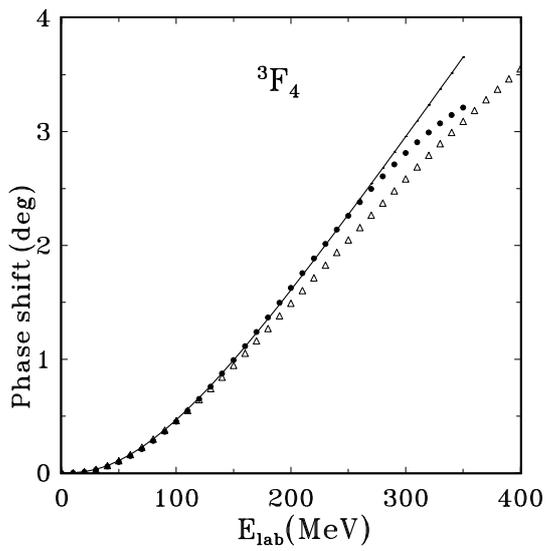
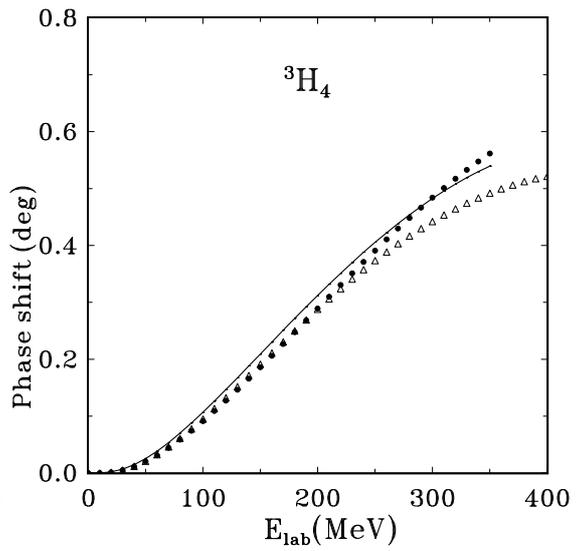
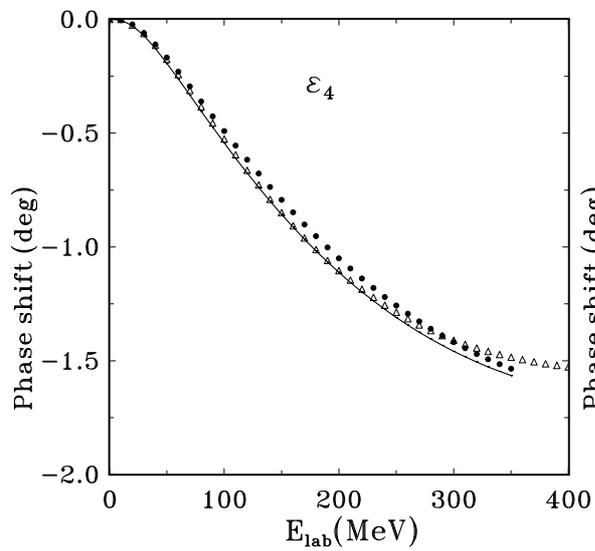
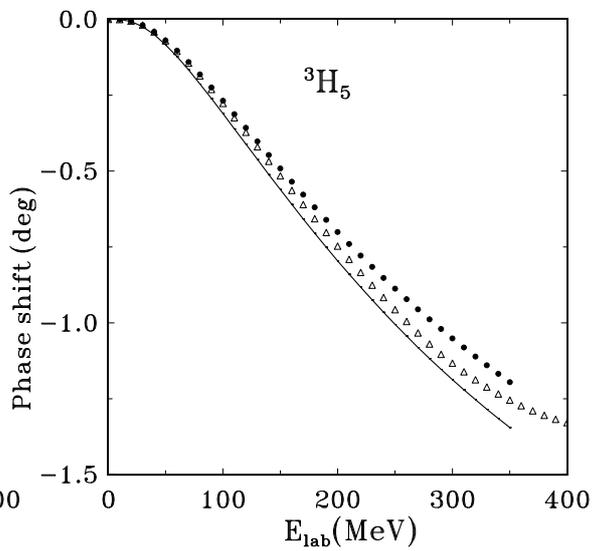
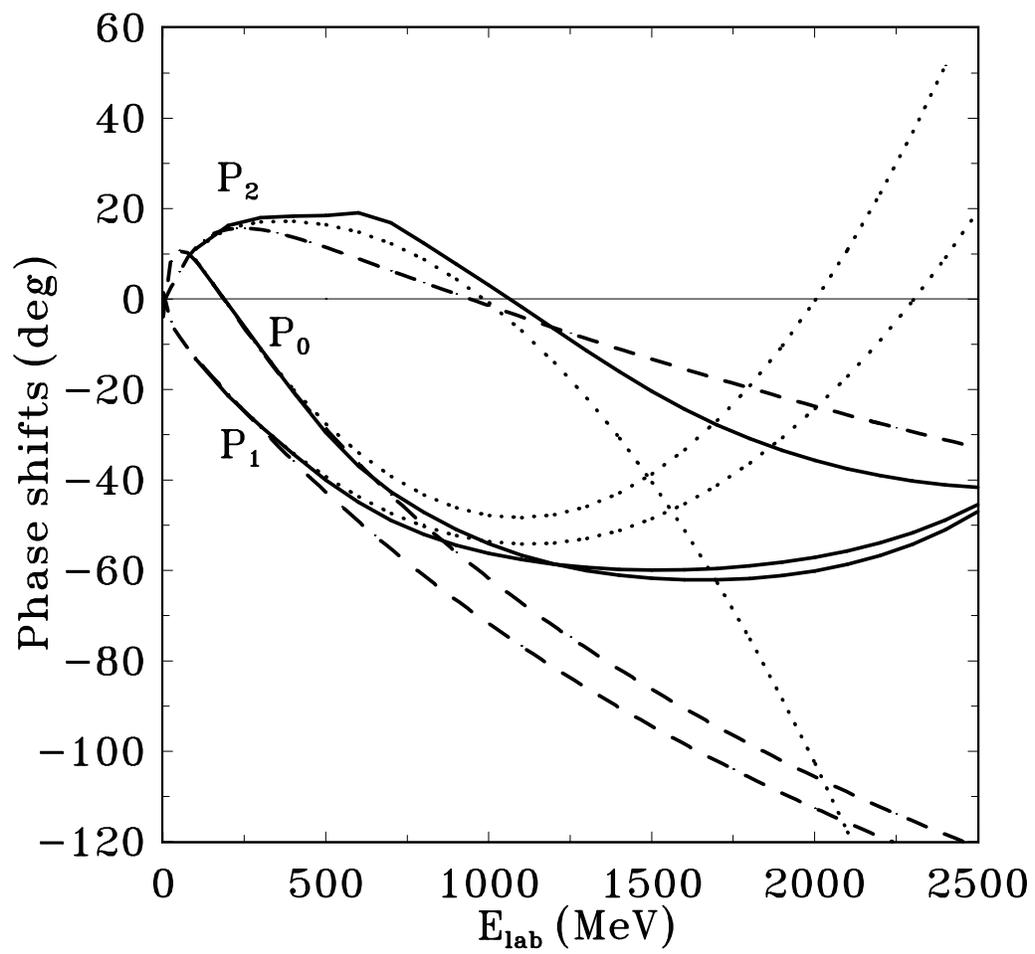
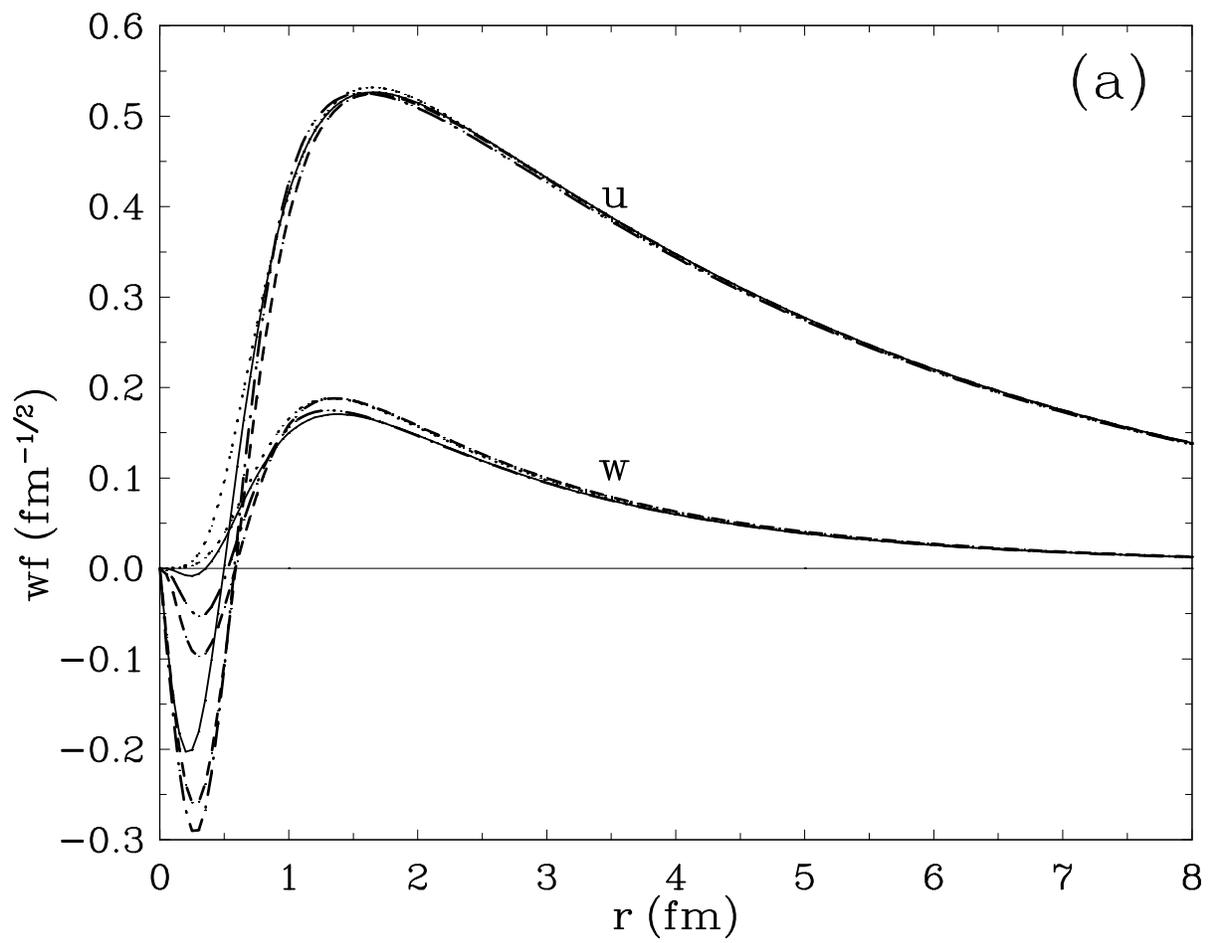
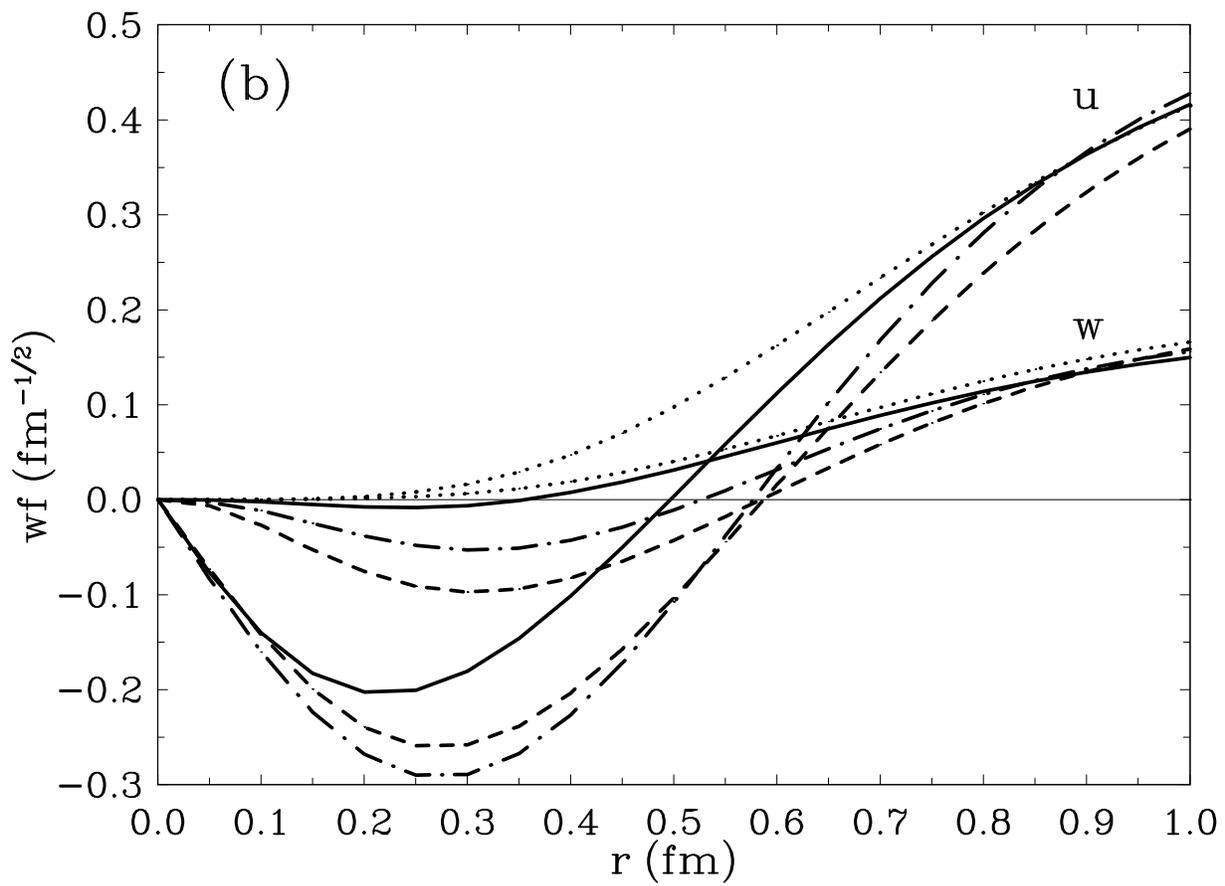
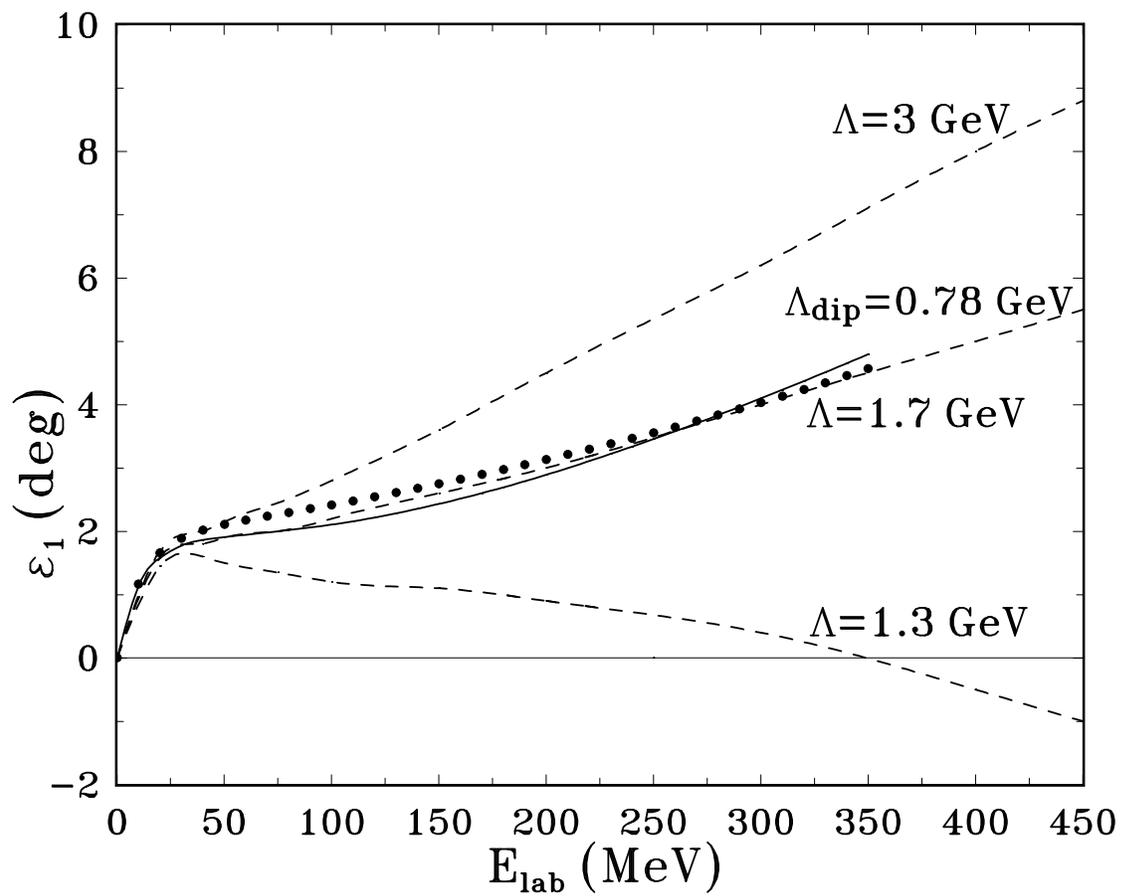
\end{document}